\documentstyle[12pt]{article}

\author{}
\date{}
\title{Caldirola-Kanai Oscillator in Classical Formulation of 
Quantum Mechanics}

\begin{document}

\maketitle

\begin{center}
{\bf S. S. Safonov}\\ 
{\it Moscow Institute of Physics and Technology}\\
{\it Institutskii pr. 9, Dolgoprudny, Moscow Region 141700, Russia} 
\end{center}

\begin{abstract}
The quadrature distribution for the quantum damped oscillator
is introduced in the framework of the formulation of quantum mechanics 
based on the tomography scheme. The probability distribution for the 
coherent and Fock states of the damped oscillator is expressed explicitly 
in terms of Gaussian and Hermite polynomials, correspondingly.
\end{abstract}

\vspace*{1cm}

In classical mechanics, the description of the motion with friction is 
described by the equation of motion has no ambiguities which are present in 
the quantum description. The quantum friction in the classical formulation of 
quantum mechanics was considered in~\cite{Safonov1}. The aim of this work is 
to discuss the problem of friction for the quantum Caldirola-Kanai 
oscillator~\cite{Caldirola,Kanai}.

Moyal \cite{Moyal} obtained a evolution equation for quantum states in
the form of the classical stochastic equation for the function which turned 
out to be the Wigner quasidistribution function~\cite{Wigner} which can not
be considered as probability since it takes negative values. Mancini,
Man'ko and Tombesi~\cite{Mancini!Man'ko} obtained the evolution equation 
for quantum state in the form of the classical stochastic equation for the 
function which turned out to be probability distribution for position
measured in ensemble of squeezed and rotated reference frames in 
the classical phase space of system. The idea of this classical-like
formulation of quantum dynamics uses the notion of optical tomography
suggested by Vogel and Risken~\cite{Vogel}. Man'ko obtained 
\cite{V.I.Man'ko} the equation for energy levels in the framework of the 
classical-like formulation of quantum mechanics and rederived the 
energy spectrum of quantum oscillator (see also~\cite{O.V.Man'ko}).

The distribution $w\left( X,\mu ,\nu ,t\right) $ for the generic linear 
combination of quadratures, which is a measurable observable 
\begin{equation}
\widehat{X}=\mu \widehat{q}+\nu \widehat{p},
\end{equation}
where $\widehat{q}$ and $\widehat{p}$ are the position and momentum,
respectively, depending on two extra real parameters $\mu $, $\nu $ is 
related to the state of the quantum system is expressed in terms of its 
Wigner function $W\left( q,p,t\right) $ as 
follows \cite{Mancini!Man'ko,V.I.Man'ko}
\begin{equation}
w\left( X,\mu ,\nu ,t\right) =\int \exp \left[ -ik\left( X-\mu q-\nu
p\right) \right] W\left( q,p,t\right) \frac{dk\, dq\, dp}{(2\pi )^2}.
\label{mar!distrib1}
\end{equation}
The distribution is normalized
\begin{equation}
\int w\left( X,\mu ,\nu ,t\right) d\, X=1.
\end{equation}
As it was shown in \cite{Caldirola,Kanai}, the quantum friction appears 
system with the Hamiltonian (we assume $\hbar =m=1$) 
\begin{equation}
\widehat{H}\left( t\right) =\frac{\widehat{p}^2}2\exp \left( -2\gamma
t\right) +\omega ^2\exp \left( 2\gamma t\right) \frac{\widehat{q}^2}2,
\label{hamiltonian}
\end{equation}
where the friction coefficient $\gamma $ and the frequency of the quantum
oscillator $\omega $ are taken to be constant. For this system, the wave 
functions of the coherent $\mid \alpha \rangle $ and Fock $\mid n\rangle $ 
states can be written as \cite{Dodonov!Man'ko:Phys.Rev.} 
(we assume $\omega =1$) 
\begin{equation}
\Psi _\alpha \left( q,t\right) =\frac 1{\sqrt[4]{\pi }\sqrt{\varepsilon }%
}\exp \left( \frac{i\stackrel{.}{\varepsilon }e^{2\gamma t}}{2\varepsilon }%
q^2+\frac{\sqrt{2}\alpha }\varepsilon q-\frac{\stackrel{.}{\varepsilon }^{*}%
}{2\varepsilon }\alpha ^2-\frac{\mid \alpha \mid ^2}2\right) ,
\label{w!fun!coher}
\end{equation}
\begin{equation}
\Psi _n\left( q,t\right) =\frac 1{\sqrt[4]{\pi }\sqrt{\varepsilon }}\left( 
\frac{\varepsilon ^{*}}{2\varepsilon }\right) ^{\frac n2}\frac 1{\sqrt{n!}%
}\exp \left( \frac{i\stackrel{.}{\varepsilon }e^{2\gamma t}}{2\varepsilon }%
q^2\right) H_n\left( \frac q{\sqrt{\varepsilon \varepsilon ^{*}}}\right) .
\label{w!fun!Fock}
\end{equation}
In these formulas the time-dependent function $\varepsilon \left( t\right) $
satisfies the equation 
\begin{equation}
\stackrel{..}{\varepsilon }\left( t\right) +2\gamma \stackrel{.}{\varepsilon 
}\left( t\right) +\varepsilon \left( t\right) =0  
\label{motion!eq}
\end{equation}
and the initial conditions 
\begin{equation}
\varepsilon \left( 0\right) =\frac 1{\sqrt{\Omega }},\ \ \ \ \stackrel{.}{%
\varepsilon }\left( 0\right) =\frac{i\Omega -\gamma }{\sqrt{\Omega }},
\label{init!cond}
\end{equation}
where $\Omega ^2=1-\gamma ^2$.
The solution $\varepsilon \left( t\right) $ has the form 
\begin{equation}
\varepsilon \left( t\right) =\frac 1{\sqrt{\Omega }}e^{-\gamma t}\left[ \cos
\left( \Omega t\right) +i\sin \left( \Omega t\right) \right] .
\label{solution}
\end{equation}

The physical meaning of the Fock state of the Caldirola-Kanai oscillator 
(\ref{w!fun!Fock}) was discussed in \cite{V.I.Man'ko}. It was shown that this 
state is a loss-energy state, and the wave function of this state has the 
property of periodicity in time with purely imaginary period.
Using the known expression of Wigner function in terms of the wave function of 
the coherent state (\ref{w!fun!coher}) (see \cite{Mancini!Man'ko,V.I.Man'ko})
and calculating the
integral (\ref{mar!distrib1}), we obtain the probability distribution for 
the coherent state 
\begin{eqnarray}
w_\alpha  &=&\frac 1{\sqrt{\pi \varepsilon \varepsilon ^{*}\left(
a^2+b^2\right) }}\exp \left( -\mid \alpha \mid ^2\right) \exp \left( -\frac{%
X^2}{\varepsilon \varepsilon ^{*}\left( a^2+b^2\right) }\right)   \nonumber
\\
&&\ \ \ \ \ \otimes \exp \left[ -\alpha ^2\frac{\varepsilon ^{*2}\left(
a-ib\right) ^2}{2\varepsilon \varepsilon ^{*}\left( a^2+b^2\right) }+\alpha 
\frac{\sqrt{2}\varepsilon ^{*}X\left( a-ib\right) }{\varepsilon \varepsilon
^{*}\left( a^2+b^2\right) }\right]   \nonumber \\
&&\ \ \ \ \ \otimes \exp \left[ -\alpha ^{*2}\frac{\varepsilon ^2\left(
a+ib\right) ^2}{2\varepsilon \varepsilon ^{*}\left( a^2+b^2\right) }+\alpha
^{*}\frac{\sqrt{2}\varepsilon X\left( a+ib\right) }{\varepsilon \varepsilon
^{*}\left( a^2+b^2\right) }\right] .  
\label{mar!distrib!coher}
\end{eqnarray}
Analogously using the wave function (\ref{w!fun!Fock}), 
we find the probability distribution for the Fock state 
\begin{equation}
w_n\left( X,\mu ,\nu ,t\right) =w_0\left( X,\mu ,\nu ,t\right) \frac
1{2^nn!}H_n^2\left( \frac X{\sqrt{\varepsilon \varepsilon ^{*}\left(
a^2+b^2\right) }}\right) ,  
\label{mar!distrib!Fock}
\end{equation}
where the probability distribution of the oscillator ground-like state is 
\begin{equation}
w_0\left( X,\mu ,\nu ,t\right) =\frac 1{\sqrt{\pi \varepsilon \varepsilon
^{*}\left( a^2+b^2\right) }}\exp \left( -\frac{X^2}{\varepsilon \varepsilon
^{*}\left( a^2+b^2\right) }\right) 
\label{mar!distrib!ground}
\end{equation}
and 
\begin{equation}
a=\frac{\exp \left( 2\gamma t\right) \nu \left( \varepsilon ^{*}\stackrel{.}{%
\varepsilon }+\varepsilon \stackrel{.}{\varepsilon ^{*}}\right) }{%
2\varepsilon \varepsilon ^{*}}+\mu ,\ \ \ \ \ 
b=\frac \nu {\varepsilon \varepsilon ^{*}}.
\end{equation}
Here $\varepsilon \left( t\right) $ is given by equation (\ref{solution}).
In Fig.~1 we show the probability distribution for the first 
excited state (loss-energy state) $w_1\left( X,\varphi ,t\right) $ as a
function of the rotation angle $\varphi $ (abscissa) and the homodyne output 
variable $X$ (ordinate)~\cite{Vogel}
\begin{equation}
\widehat{X}\left( \varphi \right) =\widehat{q}\cos \varphi -
\widehat{p}\sin \varphi .
\end{equation}  
In Fig.~1 we assume $t=5$ and $\gamma =0.05$.
 
It was shown in \cite{Mancini!Man'ko} that for the system with Hamiltonian 
\begin{equation}
\widehat{H}(t)=\frac{\widehat{p}^2}2+\widehat{V}(q,t)
\end{equation}
the quantum evolution equation alternative to the time-dependent 
Schr\"odinger equation has the form 
\begin{equation}
\stackrel{.}{w}-\mu \frac \partial {\partial \nu }w-i\left[ V\left( -\frac
1{\partial /\partial X}\frac \partial {\partial \mu }-i\frac \nu 2\frac
\partial {\partial X},t\right) -V\left( -\frac 1{\partial /\partial X}\frac
\partial {\partial \mu }+i\frac \nu 2\frac \partial {\partial X},t\right)
\right] w=0.
\end{equation}
For the damped oscillator this equation takes the form \cite{Safonov1}
\begin{equation}
\stackrel{.}{w}-\mu \frac \partial {\partial \nu }w-i\left[ \widetilde{V}%
\left( -\frac 1{\partial /\partial X}\frac \partial {\partial \mu }-i\frac
\nu 2\frac \partial {\partial X},t^{\prime }\right) -\widetilde{V}\left(
-\frac 1{\partial /\partial X}\frac \partial {\partial \mu }+i\frac \nu
2\frac \partial {\partial X},t^{\prime }\right) \right] w=0,
\label{alter!Schrod1}
\end{equation}
where
\begin{equation}
\widetilde{V}\left( q,t^{\prime }\right) =\exp \left[ 2\gamma t\left(
t^{\prime }\right) \right] V\left[ q,t\left( t^{\prime }\right) \right]
=\exp \left[ 4\gamma t\left( t^{\prime }\right) \right] \frac{q^2}2,
\label{disturb}
\end{equation}
\begin{equation}
t^{\prime }\left( t\right) =\frac{1-\exp (-2\gamma t)}{2\gamma },\ \ \ \ \
t\left( t^{\prime}\right) =-\frac{\ln (1-2\gamma t^{\prime })}{2\gamma }  
\label{t}
\end{equation}
and 
\begin{equation}
\frac{\partial t\left( t^{\prime}\right) }{\partial t^{\prime }}=\exp 
\left( 2\gamma t\right) .
\label{dt'}
\end{equation}
The dot means partial derivative in $t^{\prime }$. Using the relation 
(\ref{disturb}), one can rewrite (\ref{alter!Schrod1}) as 
\begin{equation}
\frac \partial {\partial t^{^{\prime }}}w-\mu \frac \partial {\partial \nu
}w+\exp \left( 4\gamma t\right) \nu \frac \partial {\partial \mu }w=0.
\label{alter!Schrod2}
\end{equation}
One can check that the probability distributions $w_\alpha $ 
(\ref{mar!distrib!coher}) and $w_n$ (\ref{mar!distrib!Fock}) satisfy this 
equation.

Let us consider invariants of the damped quantum oscillator 
$\widehat{a}^\dagger \widehat{a}\left( t\right) $, 
$\left( \widehat{a}^\dagger \widehat{a}\right) ^*\left( t\right) $ in the 
classical
formulation of quantum mechanics. Here asterisk means the complex conjugate 
operator. The operator $\widehat{a}^\dagger \widehat{a}\left( t\right) $
acts on the variable $q$, and the operator $\left( \widehat{a}^\dagger 
\widehat{a}\right) ^*\left( t\right) $ acts on the variable $q^\prime $ of the
density matrix $\rho _n\left( q,q^\prime ,t\right) $, which describes the Fock 
state $\mid n\rangle $ of the system. These invariants act on the distribution 
$w_n$ of the Fock state (\ref{mar!distrib!Fock}) as 
\begin{equation}
\widehat{a}^{\dagger }\widehat{a}\left( t\right) w_n\left( X,\mu ,\nu ,t\right) =nw_n\left( X,\mu ,\nu 
,t\right),  
\label{equat1b}
\end{equation}
\begin{equation}
\left( \widehat{a}^{\dagger }\widehat{a}\right) ^{*}\left( t\right) w_n\left( 
X,\mu ,\nu ,t\right) =nw_n\left( X,\mu ,\nu ,t\right) .  
\label{equat2b}
\end{equation}
The invariants $\widehat{a}^\dagger \widehat{a}\left( t\right) $ and 
$\left( \widehat{a}^\dagger \widehat{a}\right) ^*\left( t\right) $ have the 
following form
\begin{eqnarray}
a^{\dagger }a\left( t\right)  &=&\frac 12\Biggl\{ \left( \frac \partial
{\partial X}\right) ^{-2}\left[ \varepsilon \varepsilon ^{*}\left( \frac
\partial {\partial \nu }\right) ^2+\stackrel{.}{\varepsilon }\stackrel{.}{%
\varepsilon }^{*}e^{4\gamma t}\left( \frac \partial {\partial \mu }\right)
^2-e^{2\gamma t}\left( \varepsilon ^{*}\stackrel{.}{\varepsilon }%
+\varepsilon \stackrel{.}{\varepsilon }^{*}\right) \frac{\partial ^2}{%
\partial \mu \partial \nu }\right]   \nonumber \\
&&-\left( \frac \partial {\partial X}\right) ^2\left[ \varepsilon
\varepsilon ^{*}\mu ^2+\stackrel{.}{\varepsilon }\stackrel{.}{\varepsilon }%
^{*}e^{4\gamma t}\nu ^2+e^{2\gamma t}\left( \varepsilon ^{*}\stackrel{.}{%
\varepsilon }+\varepsilon \stackrel{.}{\varepsilon }^{*}\right) \mu \nu
\right]   \nonumber \\
&&+i\left[ \frac{\varepsilon \varepsilon ^{*}}2\left( \mu \frac \partial
{\partial \nu }+\frac \partial {\partial \nu }\mu \right) +\frac{\stackrel{.%
}{\varepsilon }^{*}\varepsilon e^{2\gamma t}}2\nu \frac \partial {\partial
\nu }+\frac{\varepsilon ^{*}\stackrel{.}{\varepsilon }e^{2\gamma t}}2\frac
\partial {\partial \nu }\nu \right]   \nonumber \\
&&-i\left[ \frac{\stackrel{.}{\varepsilon }\stackrel{.}{\varepsilon }^{*}}%
2\left( \nu \frac \partial {\partial \mu }+\frac \partial {\partial \mu }\nu
\right) +\frac{\varepsilon ^{*}\stackrel{.}{\varepsilon }e^{2\gamma t}}2\mu
\frac \partial {\partial \mu }+\frac{\stackrel{.}{\varepsilon }%
^{*}\varepsilon e^{2\gamma t}}2\frac \partial {\partial \mu }\mu \right] %
\Biggr\}  \label{number1}
\end{eqnarray}
and
\begin{eqnarray}
\left( a^{\dagger }a\right) ^{*}\left( t\right)  &=&\frac 12\Biggl\{ \left(
\frac \partial {\partial X}\right) ^{-2}\left[ \varepsilon \varepsilon
^{*}\left( \frac \partial {\partial \nu }\right) ^2+\stackrel{.}{\varepsilon 
}\stackrel{.}{\varepsilon }^{*}e^{4\gamma t}\left( \frac \partial {\partial
\mu }\right) ^2+e^{2\gamma t}\left( \varepsilon ^{*}\stackrel{.}{\varepsilon 
}+\varepsilon \stackrel{.}{\varepsilon }^{*}\right) \frac{\partial ^2}{%
\partial \mu \partial \nu }\right]   \nonumber \\
&&-\left( \frac \partial {\partial X}\right) ^2\left[ \varepsilon
\varepsilon ^{*}\mu ^2+\stackrel{.}{\varepsilon }\stackrel{.}{\varepsilon }%
^{*}e^{4\gamma t}\nu ^2-e^{2\gamma t}\left( \varepsilon ^{*}\stackrel{.}{%
\varepsilon }+\varepsilon \stackrel{.}{\varepsilon }^{*}\right) \mu \nu
\right]   \nonumber \\
&&-i\left[ \frac{\varepsilon \varepsilon ^{*}}2\left( \mu \frac \partial
{\partial \nu }+\frac \partial {\partial \nu }\mu \right) -\frac{\varepsilon
^{*}\stackrel{.}{\varepsilon }e^{2\gamma t}}2\nu \frac \partial {\partial
\nu }-\frac{\stackrel{.}{\varepsilon }^{*}\varepsilon e^{2\gamma t}}2\frac
\partial {\partial \nu }\nu \right]   \nonumber \\
&&+i\left[ \frac{\stackrel{.}{\varepsilon }\stackrel{.}{\varepsilon }^{*}}%
2\left( \nu \frac \partial {\partial \mu }+\frac \partial {\partial \mu }\nu
\right) -\frac{\stackrel{.}{\varepsilon }^{*}\varepsilon e^{2\gamma t}}2\mu
\frac \partial {\partial \mu }-\frac{\varepsilon ^{*}\stackrel{.}{%
\varepsilon }e^{2\gamma t}}2\frac \partial {\partial \mu }\mu \right] %
\Biggr\}.  \label{number2}
\end{eqnarray}
To obtain this form of the operators under discussion, we used the
correspondence of the action of the operators on the Wigner function 
$W\left( q,p,t\right) $ and the probability distribution 
$w\left( X,\mu ,\nu ,t\right) $~\cite{V.I.Man'ko}.

The main result of this work is the introduction of a positive normalized 
distribution function (probability distribution) for the description of the 
quantum states of the damped quantum oscillator. This distribution contains 
complete information about the state of system. For the probability 
distribution of the damped oscillator the quantum evolution equation is found, 
which is an alternative to the Schr\" odinger equation.

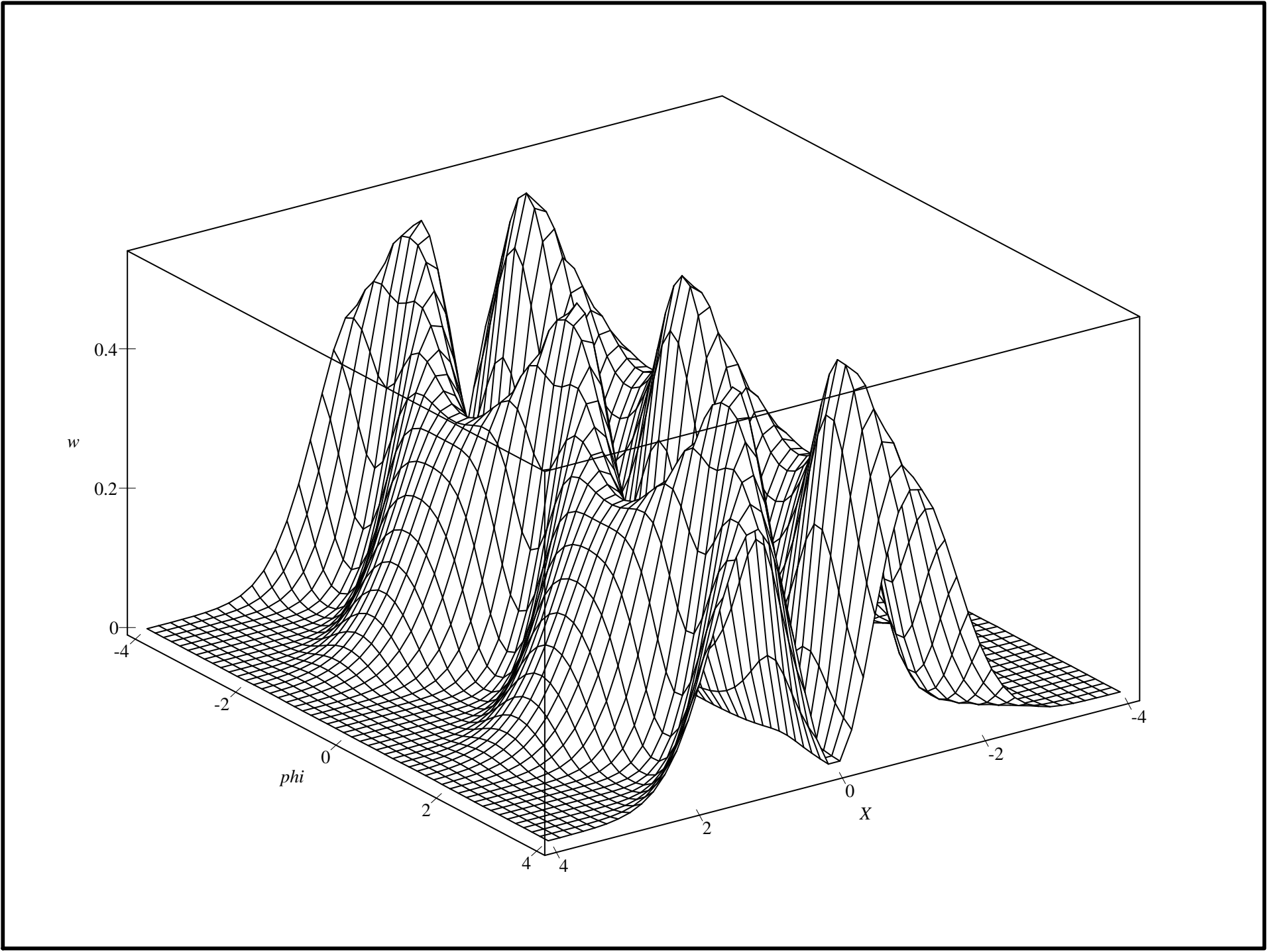
\end{document}